\documentclass[aps,
prfluids,
final
singlecolumn,
singlespace,
groupedaddress,
notitlepage 
]{revtex4-1}
\usepackage{graphicx}  
\usepackage{dcolumn}   
\usepackage{bm}        
\usepackage{amssymb}   
\usepackage{soul}
\usepackage{xcolor}
\usepackage{amsmath}
\usepackage[normalem]{ulem}
\usepackage{tikz}
\usetikzlibrary{shapes}
\usepackage[%
  colorlinks=true,%
  citecolor=blue,
  urlcolor=blue%
]{hyperref}
\usepackage{natbib}

\def \Qa {Q^{\ast}}
\def \Ma {M_{\ast}}
\def \Pe {Pe_{\ast}}
\def \rdim {\bar{r}_0}

\def \bup {\bu^{\prime}}
\def \bub {\bu_{b}}
\def \cb {c_b}
\def \cp {c^{\prime}}
\def \psib {\psi_b}
\def \psip {\psi^{\prime}}
\def \r {r_0}
\newcommand{\MU}{M = \ln(\mm/\ml)}
\def \mm   {\mu_m}
\def \ml   {\mu_l}
\def \tf   {t_f}

\def \bcdot {\boldsymbol{\cdot}}

\def \bx {\boldsymbol{x}}

\def \bu {\boldsymbol{u}}

\def \pa {\partial}

\def \wt {\widetilde}

\def \s0 {\s_0}
\def \s {\wt{\sigma}}

\def \bseq {\begin{subequations}}
\def \eseq {\end{subequations}}

\def \lap {\nabla^2} 
\def \bseq {\begin{subequation}}
\def \eseq {\end{subequation}}
\def \beq {\begin{equation}}
\def \eeq {\end{equation}}
\def \beqn {\begin{eqnarray}}
\def \eeqn {\end{eqnarray}}
\def \bi {\begin{itemize}}
\def \ei {\end{itemize}}
\def \be {\begin{enumerate}}
\def \ee {\end{enumerate}}
\def \bfig {\begin{figure}}
\def \efig {\end{figure}}
\def \ba {\begin{align}}
\def \ea {\end{align}}

\def \bseq {\begin{subequations}}
\def \eseq {\end{subequations}}

\def \bnabla {\boldsymbol{\nabla}}

\newcommand{\sqbra}[1]{\left[ #1 \right]}

\newcommand{\grad}[1]{\boldsymbol{\nabla} #1}
\newcommand{\divg}[1]{\bnabla \bcdot #1}

\newcommand{\modu}[1]{\left\lvert #1 \right\rvert}

\begin{document}



\title{Control of radial miscible viscous fingering} 
\author{Vandita Sharma}
\affiliation{Department of Mathematics, Indian Institute of Technology Ropar, Rupnagar - 140001, India}
\author{Sada Nand}
\affiliation{Department of Mathematics, Indian Institute of Technology Ropar, Rupnagar - 140001, India}
\author{Satyajit Pramanik}
\affiliation{NORDITA, Royal Institute of Technology and Stockholm University, 106 91  Stockholm, Sweden}
\author{Ching-Yao Chen}
\affiliation{Department of Mechanical Engineering, National Chiao Tung University, Hsinchu, Taiwan, 30010 Republic of China}
\author{Manoranjan Mishra}
\affiliation{Department of Mathematics, Indian Institute of Technology Ropar, Rupnagar - 140001, India}
\affiliation{Department of Chemical Engineering, Indian Institute of Technology Ropar, Rupnagar - 140001, India}
\date{\today}

\begin{abstract}
We investigate the stability of radial viscous fingering (VF) in miscible fluids. We show that the instability is decided by an interplay between advection and diffusion during initial stages of flow. Using linear stability analysis and nonlinear simulations, we demonstrate that this competition is a function of the radius $\r$ of the circular region initially occupied by the less viscous fluid in the porous medium.  For each $\r$, we further determine the stability in terms of P\'eclet number ($Pe$) and log-mobility ratio ($M$). The $Pe-M$ parameter space is divided into stable and unstable zones--the boundary between the two zones is well approximated by $M =\alpha(\r) Pe^{-0.55}$. In the unstable zone, the instability is reduced (enhanced) with an increase (decrease) in $\r$. Thus, a natural control measure for miscible radial VF in terms of $\r$ is established. Finally, the results are validated by performing experiments which  provide a good qualitative agreement with our numerical study. Implications for observations in oil recovery and other fingering instabilities are discussed. 
\end{abstract}


\maketitle 
\emph{Introduction}--Hydrodynamic instabilities are ubiquitous to transport in porous media. Viscous fingering (VF) is one of these instabilities, which is observable while displacing a less mobile fluid by another more mobile fluid through porous media, and it plays critical roles in enhanced oil recovery through miscible flooding/solvent drive \cite{Lake1989enhanced}, 
chromatography separation \cite{Guiochon2008}, pattern formation \cite{Li2009}, medicines \cite{Bhaskar1992}, CO$_2$ sequestration \cite{Moortgat2016, Amooie2017_GRL,Qian2019}, diffusion-limited aggregation \cite{Witten1981}, mixing \cite{Jha2011a}, 
and bacterial colonies \cite{Callan2008}. 

While advection is necessary for VF, diffusion stabilizes this instability in miscible systems. For a given pressure gradient, the advection velocity is uniform in a rectilinear flow, whereas, in a radial flow, the velocity is inversely proportional to the radial distance from the point of fluid injection. Effects of diffusion on stabilization of miscible fingering instabilities both in the linear and nonlinear regimes have been well understood mainly in the context of rectilinear displacement flows \cite{Homsy1987, Pramanik2015b}. In radial VF, the effects of diffusion at the initial stages of the displacement have been studied by \citet{Tan1987}, who concluded that the dispersion is strong enough to completely suppress the instability when the P\'eclet number $\sim O(1)$; otherwise the displacement is always unstable.  On the other hand, \citet{Chui2015} observed shut down of the overall flow instability emerging from the dominance of diffusion over advection at the later stages of the flow. Most recently, diffusion-driven transition between two regimes of VF is captured \cite{ThomasPhysRevFluids2019}. Thus, there are many facets of the competition between advection and diffusion in radial flows. Contrary to this, \citet{Bischofberger2014} claim that viscosity ratio sets the velocity of the interface and three regimes of instability are obtained, for which the effects of diffusion are irrelevant. In this paper, we show that the diffusion can never be neglected when dealing with miscible fluids.  Initial stable displacement is the characteristic of dominant diffusion, which is equilibrated by advection at a later stage that is identified as the transition to an unstable state dominated by advection.  

Further, we ask: Can we control the competition between advection and diffusion to suppress the miscible radial VF? Diffusion, being an inherent property that depends on the displacing and displaced fluids, is difficult to tune; however, the advection can be suitably modified. Many studies focussed on controlling VF \cite{Dias2012, Zheng2015,Yuan2019}  utilise time-dependent strategies to control advection. However, we achieve the same by merely modifying the initial configuration of radial displacement flow. We consider different initial finite volume of the displacing fluid in the porous medium, which is  represented by an initial radial distance ($\r$) of the interface from the center of the porous medium. The effects of competition between advection and diffusion on the controllability of VF are parametrised in terms of $\r$  and are explained through  linear stability analysis (LSA) and compared with the corresponding non-linear simulations (NLS); supported by the result of a diligently designed experiment.  

\emph{Mathematical formulation and linear stability analysis}--The fluids considered are Newtonian, miscible, non-reactive with $\ml, \mm$ as viscosity of the less and the more viscous fluid, respectively. The non-dimensional governing equations for the flow in a two dimensional (2-D) homogeneous porous medium are constituted by the Darcy's law and the transport equation for the solute concentration $c$,
\begin{eqnarray}
\label{eq:cont_non_dim}
& & \divg{\bu} = 0, \\
\label{eq:Darcy_non_dim}
& & \bu = - \frac{1}{\mu(c)}\grad{p}, \quad \mu(c) = e^{(1-c)M}, \\ 
\label{eq:Mass_non_dim}
& & \pa_t c + \bu \bcdot \grad{c} = \frac{1}{Pe} \lap{c}, 
\end{eqnarray}
where $p$ is the hydrodynamic pressure and $\bu=(u,v)$ is the Darcy velocity vector. We use $\tf$, the total time of fluid injection, as the characteristic time, and $\sqrt{Q \tf}$ as characteristic length, where $Q$ is the gap-averaged flow rate. Consequently, we obtain two non-dimensional parameters:  P\'eclet number, $Pe=Q/D$, and the log-mobility ratio $\MU$, where the molecular diffusion coefficient $D$, of $c$ in the solvent fluid is assumed to be a constant. 

We consider a 2-D square domain $\Omega =[-1.5,1.5] \times [-1.5,1.5]$ in the cartesian coordinates with the origin as the source of the less viscous fluid [see Appendix \ref{App:Compu} for computational domain]. Initial condition associated with above equations is 
$ \bu(\bx, t=0) = \bx/(2 \pi |\bx|^2)$
and 
\beq
\label{eq:IC}
c(\bx, t = 0) = \left\{
\begin{array}{ll}
1, & 0 \leq |\bx|^2 \leq \r^2 \\ 
0, & \mbox{Otherwise} \\
\end{array} 
\right.,  
\eeq	
where $\bx=(x, y)$, and $\r$ is the non-dimensional radius of initial circle occupied by less viscous fluid. 

\begin{figure}[!htbp]
\centering
\includegraphics[trim={0 0 0 0},clip, scale=0.8]{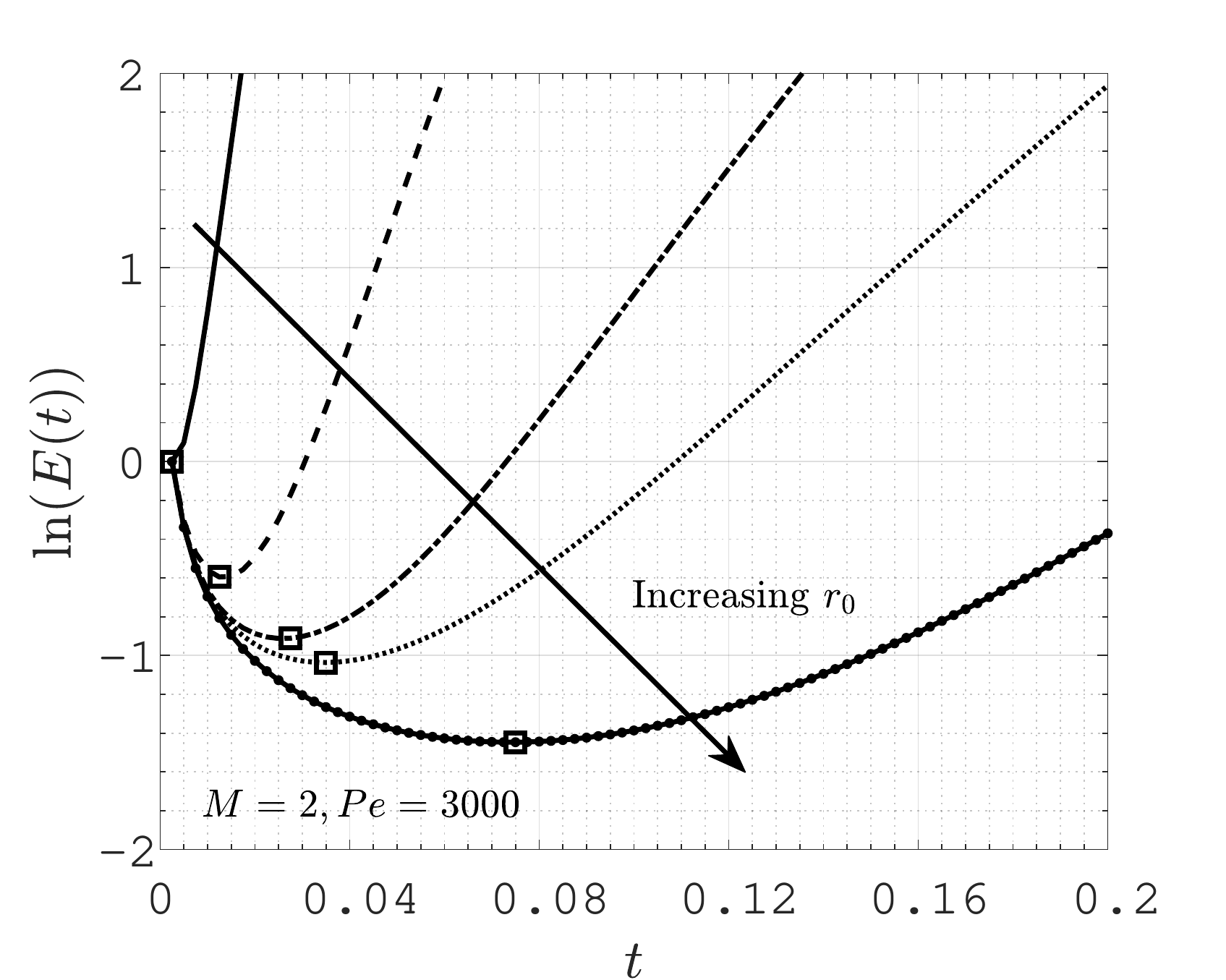}
\caption{Temporal evolution of $\ln(E(t))$ for $\r = 0.1$ to $0.3$ with an increment $0.05$. The onset time of instability is marked by $\protect \square$. The initial diffusion is prevalent for a longer time for a larger $\r$ and hence the onset is delayed.} 
\label{fig:M2_LSA}
\end{figure}

First, we perform LSA to identify the effects of diffusion at the initial stages of the displacement and the onset of VF. 
Assume the base state velocity to be $\bu_b(\bx) = \bx/(2 \pi |\bx|^2)$, and the base state concentration $\cb$ is the solution of Eq. \eqref{eq:Mass_non_dim} for the initial condition \eqref{eq:IC}. Analytical solution for this initial-boundary value problem is not attainable. We use the method of lines to numerically compute $\cb$. Spatial derivatives are discretised using sixth-order compact finite differences \cite{Lele1992} and the resulting initial value problem is solved using a third-order Runge Kutta method. The velocity field is solved in the form of stream-functions $\psi(x, y)$, defined as $u = \pa_y \psi, \; v = -\pa_x \psi$. 
We introduce an infinitesimal perturbation ($\psip, \; \cp$) such that $ \psi = \psib + \psip, c = \cb + \cp$ ($\modu{\psip}, \modu{\cp} \ll 1$), where $\psi_b$ is the stream-function corresponding to $\bu_b$. The corresponding linearized equations are solved for $\psip, \; \cp$ using the hybrid compact finite differences and the pseudo-spectral method. No flux boundary condition for $\cp$, and  $\psip = 0$ are used at the outflow boundary. The present LSA works as an alternative approach to study time-dependent linear system arising in miscible VF. Our interest does not lie in the  wavelength selection like many other LSA \cite{Hota2015b}, but  to capture initial diffusion and its effect on the onset of instability. However, it must be noted that our LSA is also applicable for wavelength selection \cite{Hota2015a}.

Recall that both $\cb$ and $\cp$ evolve temporally. Therefore, the growths of $\cp, \; \psip$ are relative to that of $\cb, \; \psi_b$, and we quantify them at each instant of time. We define the energy ratio 
$R(t) = E_K(\cp, \bup)/E_K(\cb, \bub)$, 
where 
$E_K(c, \bu) =  \int\limits_\Omega \sqbra{ c^2 + \bu^2 } d\Omega$ represents the kinetic energy at time $t$. 

\begin{figure}[!htbp]
\centering 
\includegraphics[trim={15 0 0 0},clip, scale=0.8]{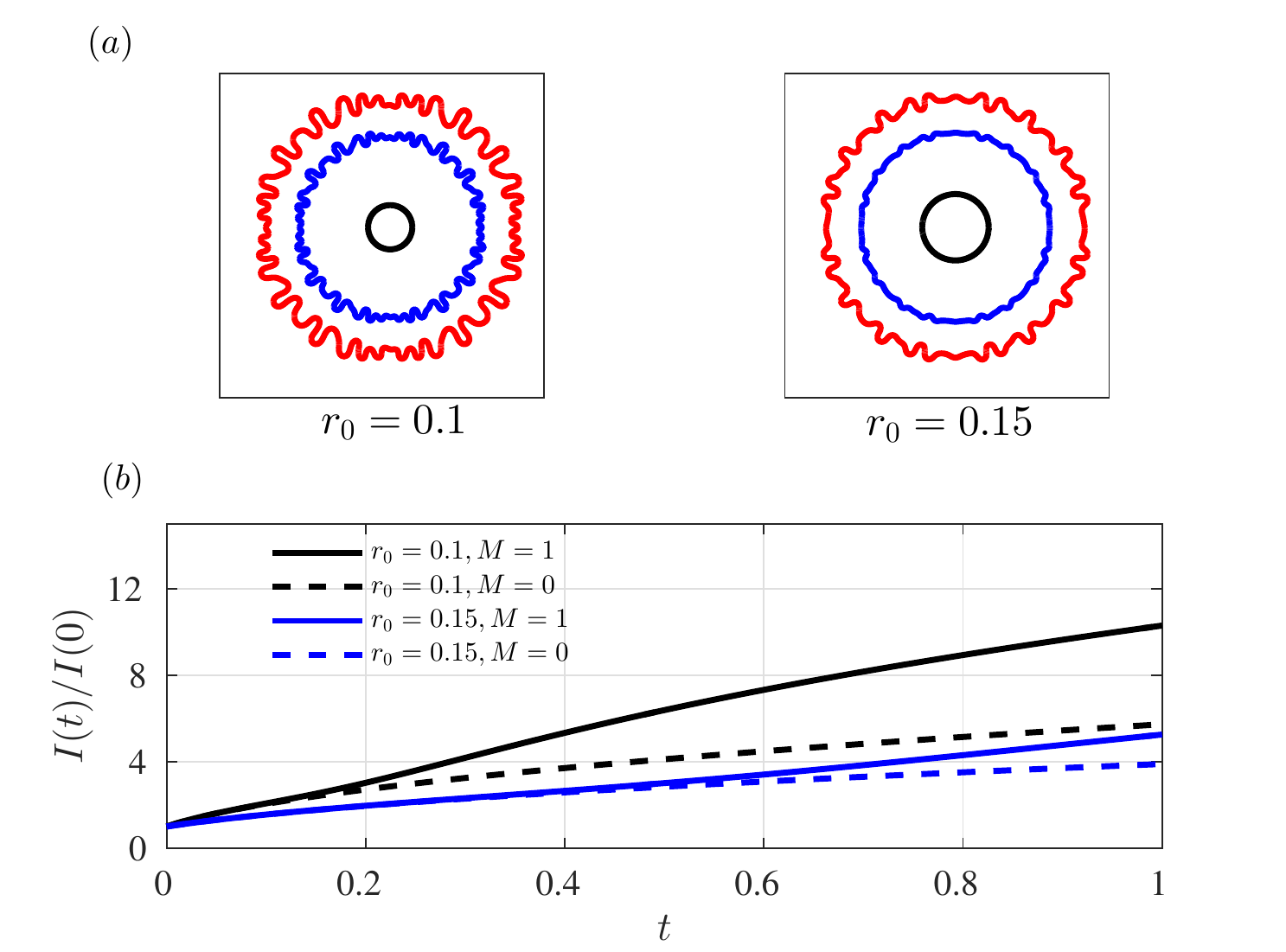}
\caption{$(a$) Concentration contour $c = 0.5$ for $Pe = 5000, \; M = 1$ at $t = 0$ (black), $0.5$ (blue), and $1$ (red). Increasing $\r$ abates the instability as evident from the less distorted contours at same time. $(b$) Temporal evolution of normalized $I(t)$. Maiden deviation of the solid line  from the dashed line marks onset of instability.
}
\label{fig:M2_Inter}
\end{figure}

The energy amplification, $E(t)= R(t)/R(t=0)$ is the ratio of the energy at time $t$ to its value at $t = 0$  \cite{Matar1999}. 
The nature of amplification as a function of time decides stability. An increasing (decreasing) $E(t)$ indicates a relative growth (decay) of the perturbations, while the presence of an extremum, if any, is of special importance. A minimum indicates transition from diffusion-dominating region to an advection-dominated one, which eventually implies the triggering of instability; while a maximum exhibits transient growths \cite{Hota2015b}. Figure \ref{fig:M2_LSA} shows the natural logarithm of energy amplification as a function of time. Evidently, $\ln(E(t))$ is a non-monotonic function with a minimum occurring for each $\r > 0.1$. The minimum captures the competition between advection and diffusion in the linear regime. Up to the point of  minimum, the disturbances are stabilized by the diffusive base state. Larger the $\r$, later the minimum is obtained, indicating a delayed onset due to the dominance of diffusive forces for a longer time. Similar qualitative results have been  verified for various $M$ and $Pe$. Therefore our LSA captures controllability of VF due to the competition between the two forces.  

\begin{figure}[!htbp]
\centering
\includegraphics[trim={0 0 0 0},clip, scale=0.8]{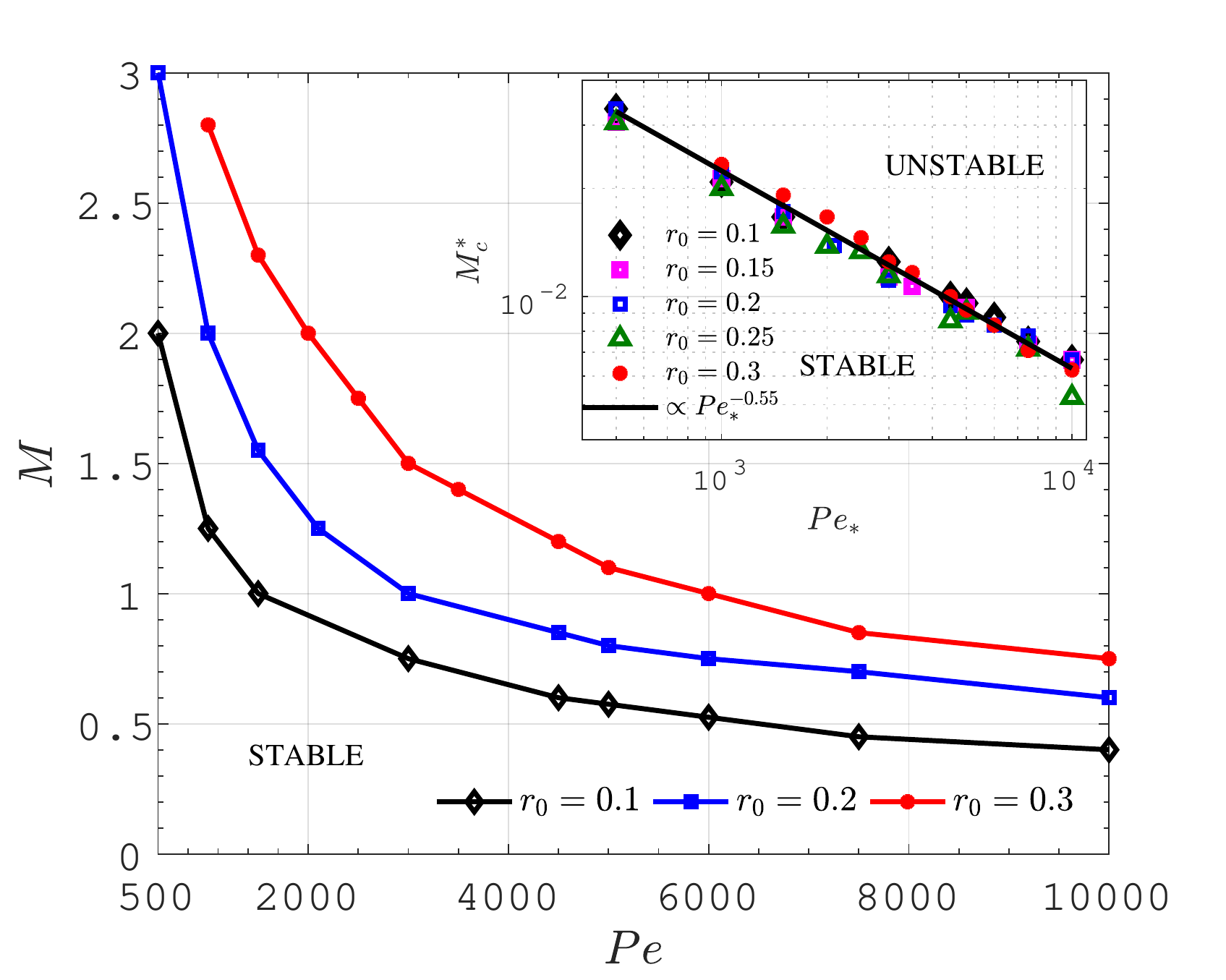}
\caption{$M-Pe$ parameter space divided into stable and unstable regions for each $\r$. The ordered pair $(Pe, \; M)$ below a curve corresponding to a given $\r$ associates a stable displacement for the corresponding $\r$. Inset: Boundary between the two zones scales as $M_c^* =  Pe_*^{-0.55},$ where $M_c^*=M_*/[30(10~\r+1)]$.} 
\label{fig:M_Pe}
\end{figure}

\emph{Nonlinear simulations}--Next, we perform NLS to support the estimates of LSA that the instability is delayed as $\r$ increases. We solve the coupled non-linear equations, Eqs. \eqref{eq:Darcy_non_dim}--\eqref{eq:IC}, using a hybrid scheme based on compact finite differences and pseudo-spectral methods. This method has been extensively used to study instabilities in porous media \cite[and Refs. therein]{Chen2010, Sharma2019}. We perform NLS for five different values of $\r$, and for each $\r$, we consider $M \in [0,3]$ and $Pe \in [500,10^4]$. A comparison of the concentration contours at a given time for the two radii in Fig. \ref{fig:M2_Inter}(a) clearly depicts that fingering instability is abated as $\r$ increases. Thus, the NLS support the instability control predicted by LSA.  For each simulation we compute the interfacial length \cite{Mishra2008}, $I(t) = \int\limits_\Omega |\grad{c}| d\Omega$, which measures the temporal variation of the concentration gradient. For $M = 0$, we find $I(t) = 2 \sqrt{\pi (t + \pi \r^2)} := I_0(t)$ (say); in the absence of VF, $I(t)$ coincides with $I_0(t)$. We define $t = t_{on}$ as onset time of fingering in the nonlinear regime if $I(t) > I_0(t), \; \forall t$ satisfying $t_{on} \lesssim t < 1$, and the corresponding parameter set ($\r, M, Pe$) is identified as \emph{unstable}, otherwise \emph{stable}. For each $\r$, we summarize the instability in the $Pe-M$ parameter space; see Fig. \ref{fig:M_Pe} for $\r = 0.1, \; 0.2$ and $0.3$. This indicates that for a fixed $M$, there is a critical P\'eclet number (or, similarly for a fixed $Pe$, there is a critical log-mobility ratio) for the occurrence of instability. The existence of a critical parameter for fingering instability in radial source flow with point source (i.e., $\r = 0$) is already identified using LSA \cite{Tan1987} and experiments \cite{Bischofberger2014,ThomasPhysRevFluids2019}. Furthermore, from NLS we observe that $t_{on}$ increases with $\r$ and the stable region spans over a larger range of both $M$ and $Pe$ [see Appendix \ref{App:NLS} for density plots of concentration]. It is noteworthy that the boundary between the stable and unstable regions follow a scaling relation $\Ma = \alpha(\r) \Pe^{-\beta}$, with $0.52 \lesssim \beta \lesssim 0.59$. We observe that the parameter pair ($\Ma, \Pe$) lying on the boundary between the stable and the unstable regions can be well approximated by the relation $\Ma = 30(10\r + 1) \Pe^{-0.55}$ [inset in Fig. \ref{fig:M_Pe}]. Therefore, using this scaling relation we can approximate the stability of radial flows in homogeneous porous media.

\begin{figure}[!htbp]
\includegraphics[trim={20 0 0 0},clip, scale=0.8]{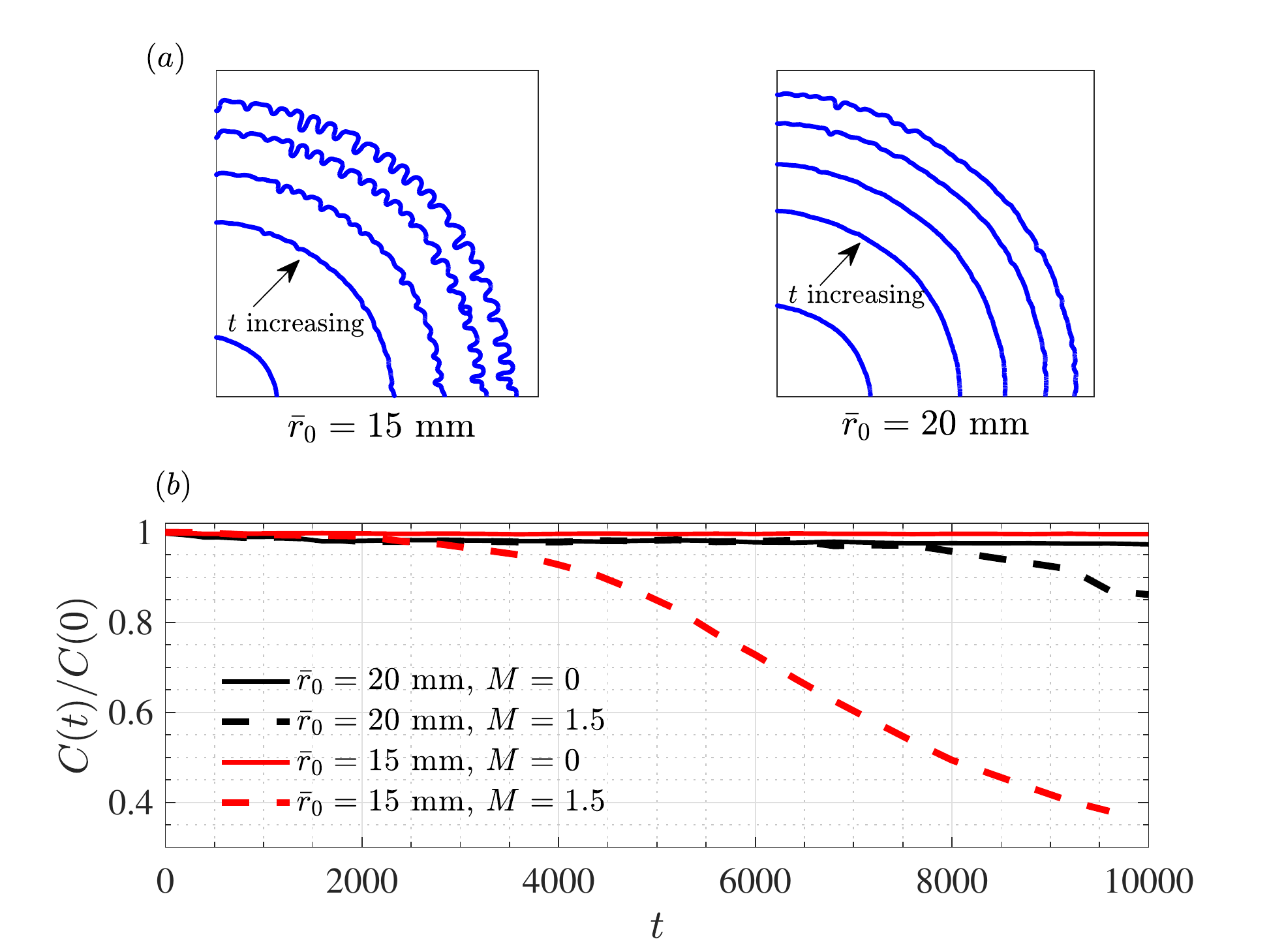}
\caption{$(a)$ Contours  at $t = 0, 4000, 6000, 8000, 9600$ ms of the experimental images showing controlling effect of finiteness in terms of reduced and delayed instability. $(b)$ Circularity as a function of time showing a delayed occurrence of distortions for larger $\rdim$.}
\label{fig:EXP}
\end{figure}

\emph{Experiments}--Our numerical results are further validated through state-of-the-art experiments. The less viscous fluid is injected at a constant flow rate $Q_1$ ml/s for a period of $t_1$ s in the Hele-Shaw cell initially filled with the more viscous fluid [see Appendix \ref{App:Expe_set}, \ref{App:Expe_IC}  for details of experimental set up]. As soon as the required dimensional radius $\rdim$, is reached, the flow rate is increased to $\Qa$ and the less viscous fluid is continuously injected at this flow for a final time fixed for all the experiments. We repeat a series of experiments with $\Qa \in [0.05, 0.5]$ ml/s, $M \in [0, 2]$, and capture suppression of fingering instability for different values of $\Qa$ and $M$ qualitatively similar to NLS. No instability is observed for many $\Qa$ suggesting that there always exists a stable zone for each $\rdim$ as predicted by NLS for a range of $Pe$ and shown in Fig. \ref{fig:M_Pe}. We use in-built MATLAB command \textit{imcontour} to plot the contours at the interface. For visualisation purpose, we show the contours of one quarter the experimental image in Fig. \ref{fig:EXP}(a). A delayed onset and reduced fingering corresponding to $\rdim = 20$ mm compared to $\rdim = 15$ mm for $\Qa = 0.4$ ml/s, and $M = 1.5$ are evident. This supports the fact that larger the $\rdim$, weaker is the advection and hence the initial competition between advection and diffusion determines the instability [see Appendix \ref{App:Expe_r0} for a comparison with $\rdim=0$]. 

We use IMAGEJ \cite{Schneider2012} to quantify the experimental results. In order to avoid the noise induced by the presence of the injection pipe\textbf{s} in the images, we utilize half of the experimental domain in our analysis. The numerical study done so far predicts stable radial displacement upto some initial time due to the dominance of diffusive forces over the advective. Hence, for the validity of the control measure, the experiments should also capture the initial radial displacement in the form of a circular displacing front. Circularity is used as the measure of the extent up to which the displacement front is circular \cite{Escala2019}. We define the  circularity as $C(t) = 2 \pi A(t)/P^2(t)$, where $A(t)$ and $P(t)$ correspond to the area and the perimeter of the region occupied by the less viscous fluid, respectively; so that $C(t) =1$ for a semi-circle. For $P(t)$, we consider only the length of the curved surface since the diameter does not contribute to $C(t)$. Subject to experimental errors, $C(t)$  close to $1$ but constant over the frame of time implies a circular displacing front. The maiden deviation of the circularity from the constant value indicates distortions at the front and it marks the triggering of the instability. $M=0$ works as the ideal radial source flow as no instability is observed as a result of equal viscosity of the two fluids. Hence, $C(t)$ for $M=0$ is used as the reference constant value for each $\rdim$. $C(t)$ for $M \neq 0$ deviates from the constant value after some initial time for each $\rdim$ and the time of deviation is larger for larger $\rdim$ (see Fig. \ref{fig:EXP}(b)). In other words, the instability is triggered later for a larger $\rdim$ [see Appendix \ref{App:Compu} also for the circularity calculated from NLS]. This reassures a qualitative agreement of the experiments with LSA and NLS. The proposed control measure can be used to control VF in majority of radial source flows with miscible fluids. 

\emph{Discussions and conclusion}--Depending upon the application, controlling fingering instabilities is of paramount importance, such as mixing can be increased by increasing the instability; whereas for an improved oil recovery or separation process, instability should be suppressed. 
VF in immiscible systems are controllable by modifying the geometry \cite{Bongrand2018, Puzovic2018}, using time dependent strategies \cite{Dias2012, Zheng2015}. Though all these control mechanisms may be directly inapplicable in miscible VF \cite{Huang2015}, in this paper, we investigate a stability mechanism in miscible VF based on the principle of stabilisation in immiscible flows. 
Using LSA and NLS we obtain a criterion to control radial miscible VF. We utilize the competition between the advective and diffusive forces to control the miscible VF in a radial flow and show that the initial position of the miscible interface (the radius of the initial circular region containing the displacing fluid) is a control parameter. No modification in the geometry or a continuous variation in the flow rate  is required unlike many other earlier studies of immiscible VF \cite{Zheng2015, Dias2012}. Our theoretical results are compared with experiments that are in qualitative agreement with the numerical results. The quantitative differences between the experimental and numerical results are attributed to the gap-wise variation of the flow in the Hele-Shaw experiments compared to the 2-D numerical simulations. 

Time-dependent strategies, such as changing the flow rate in a rectilinear configuration \cite{Yuan2019} or varying the gap-width of the Hele-Shaw cell in a radial configuration \cite{Chen2010}, are attempted numerically in the context of miscible VF. Nevertheless, these studies lack sufficient evidences about the complete suppression of instability. Moreover, these strategies are not experimentally validated. Contrary to these, the highlights of our analysis is to successfully generate a stable radial experiments in accordance to our stability analysis and numerical simulations. 

LSA predicts a delayed onset of instability with increasing $\r$, while NLS predict a critical log-mobility ratio $M_*$ upto which there is no instability, dividing the $M-Pe$ plane into stable and unstable zones for each $\r$. The log-mobility ratio and P\'eclet number on the boundary of the stable and unstable zones scale as $M_* = 30(10 \r+1) Pe_*^{-0.55}$. The stable zone increases with increasing $\r$. Taking $\r \rightarrow 0$ we approximate the critical $Pe$ for a point source radial flow, which, for a mobility ratio $3$, is $\approx 65.8$ that is of the same order as estimated from linear stability by \citet{Tan1987}. For fixed values of $Pe$ and $M$, it is concluded that a stable displacement for a given $\r$ ensures a stable displacement for all larger $\r$ despite a favorable viscosity gradient. On the other hand, an unstable displacement for a given $\r$ indicates a stronger and early instability for all smaller $\r$ keeping $M$ and $Pe$ fixed. This accounts to weakening of advection with an increase in the distance from the source. Experiments performed depict the validity and the applicability of the proposed control strategy. 

Beside helping to understand the intrinsic properties of fundamental hydrodynamic instabilities \cite{Paterson1981, Tan1987, Bischofberger2014} and pattern formation \cite{Witten1981, Li2009}, our results suggest that the controllability of miscible VF in a radial configuration could be important to predict the effectiveness of enhanced oil recovery by polymer flooding \cite{Lake1989enhanced} and in various other similar configurations. 

\section*{Acknowledgment}
M.M. acknowledges the financial support from SERB, Government of India through
project grant no. MTR/2017/000283. S.P. acknowledges the support of the Swedish
Research Council Grant no. 638-2013-9243. C.-Y.C. is thankful to ROC (Taiwan)
Ministry of Science and Technology, for financial support through Grant no. MOST
105-2221-E-009-074-MY3. V.S. acknowledges 2017 NCTU Taiwan Elite Internship Program for
the financial support to visit C.-Y.C.


%

\clearpage

\appendix

\section{Linear stability analysis and non linear simulations}

\subsection{Computational domain and circularity}
\label{App:Compu}

\begin{figure}[!htbp]
 \centering
 $(a)$ \hspace{3.2in} $(b)$ \\
\includegraphics[trim={0 40 9.5cm 50},clip, scale=0.5]{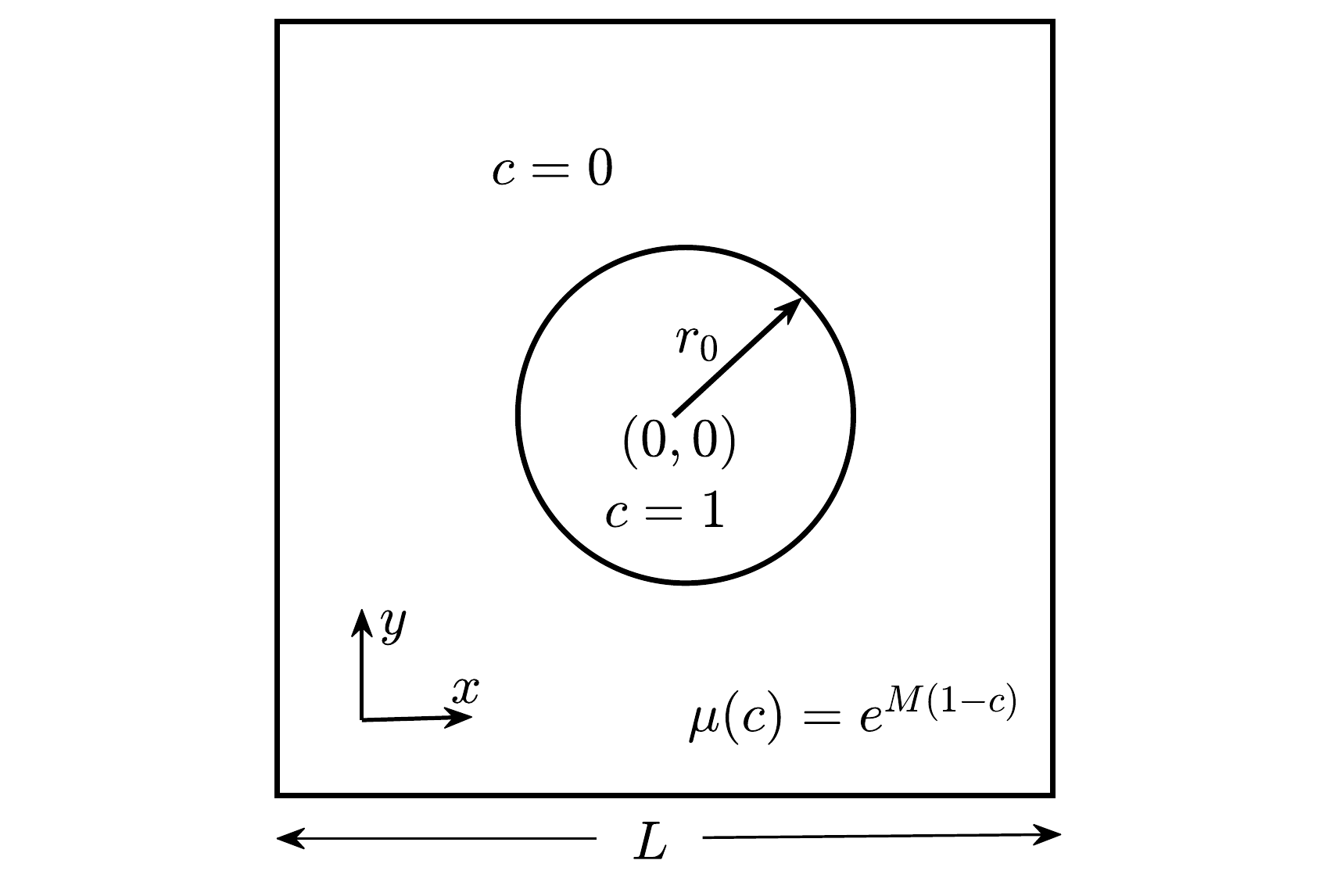}
\includegraphics[trim={0 0 0 0},clip, scale=0.55]{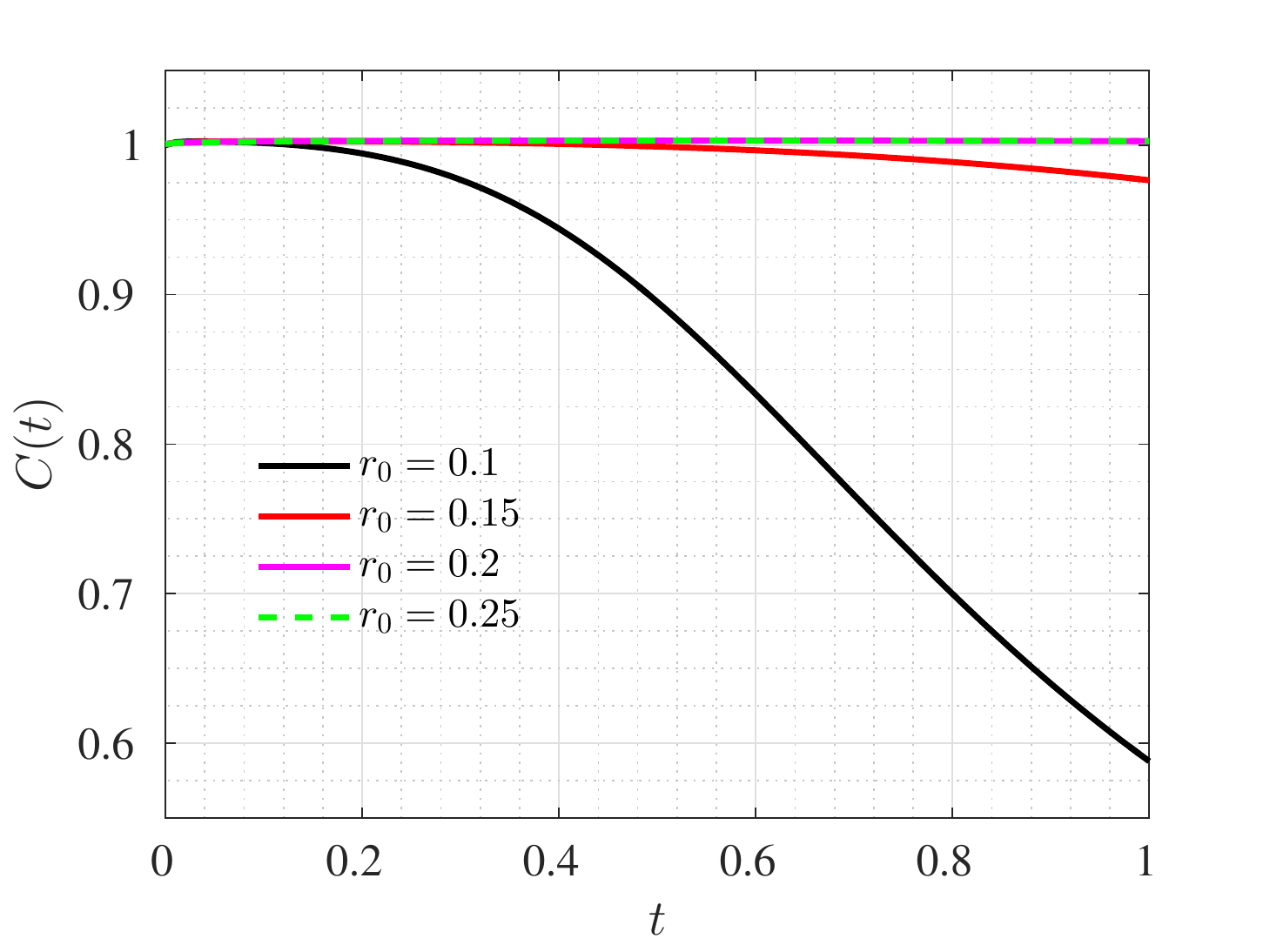}
  \caption{$(a)$  Schematic of the computational domain $ \Omega= [-L/2,L/2] \times [-L/2,L/2]$, with $L=3$ used for LSA and NLS. The center of the domain is chosen as the origin of the cartesian coordinate system and the source of the less viscous fluid initially occupying a circle of radius $\r$. Here $M>0$. $(b)$ Circularity  computed from numerical data,  $C(t) = 4 \pi A(t)/P^2(t) =4 \pi  \int\limits_\Omega c d\Omega/(\int\limits_\Omega |\grad{c}| d\Omega)^2$. A delay in onset with increasing $\r$ and a suppressed instability is captured by $C(t)$ for $M=2, Pe=1000$.} 
\label{fig:schematic}
\end{figure}

\subsection{Density plots from NLS}
\label{App:NLS}
\begin{figure}[h]
$(a)$ \hspace{3in} $(b)$ 
 \centering
\includegraphics[trim={0 0 0 0},clip, scale=0.47]{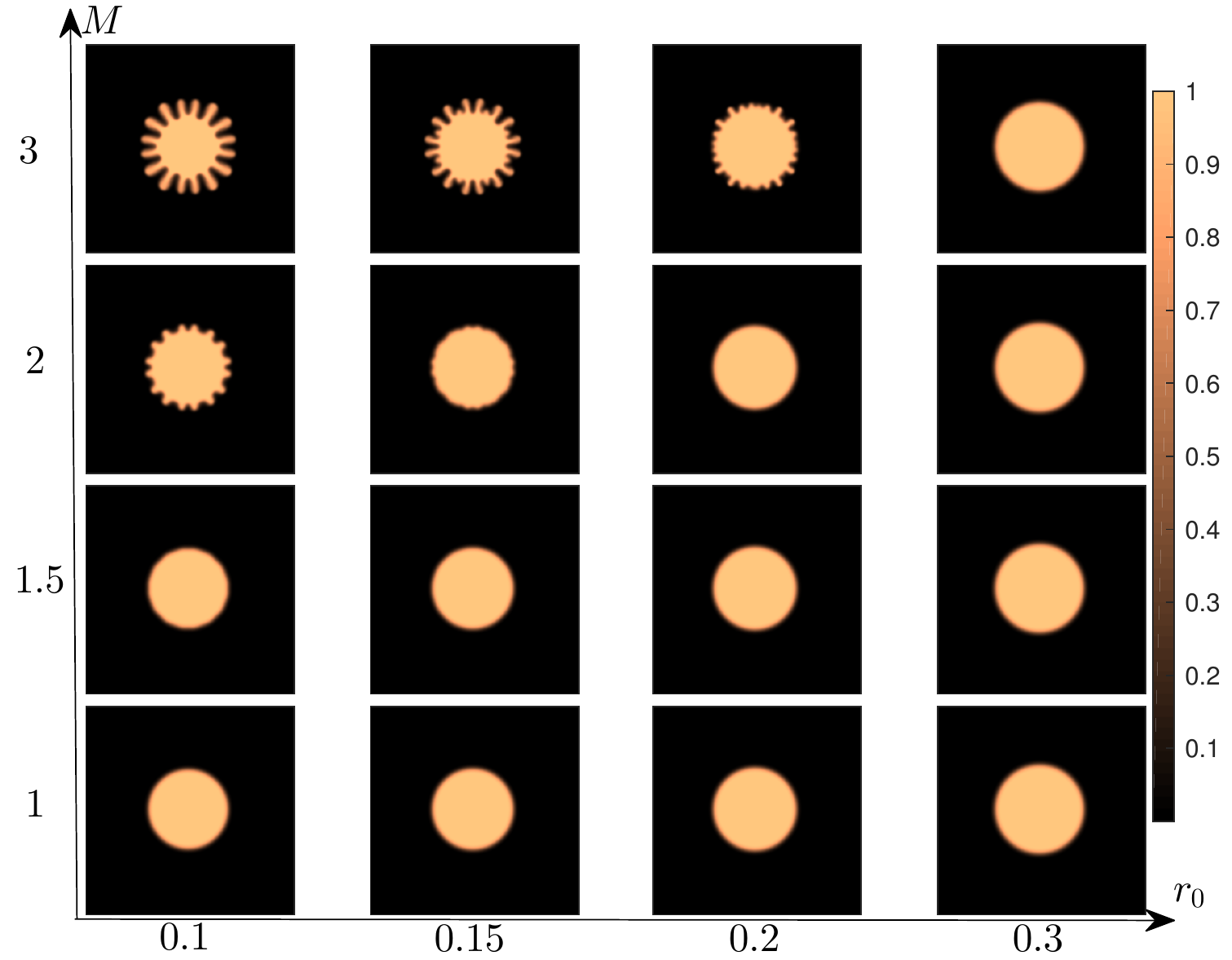}
\includegraphics[trim={0 0 0 0},clip, scale=0.5]{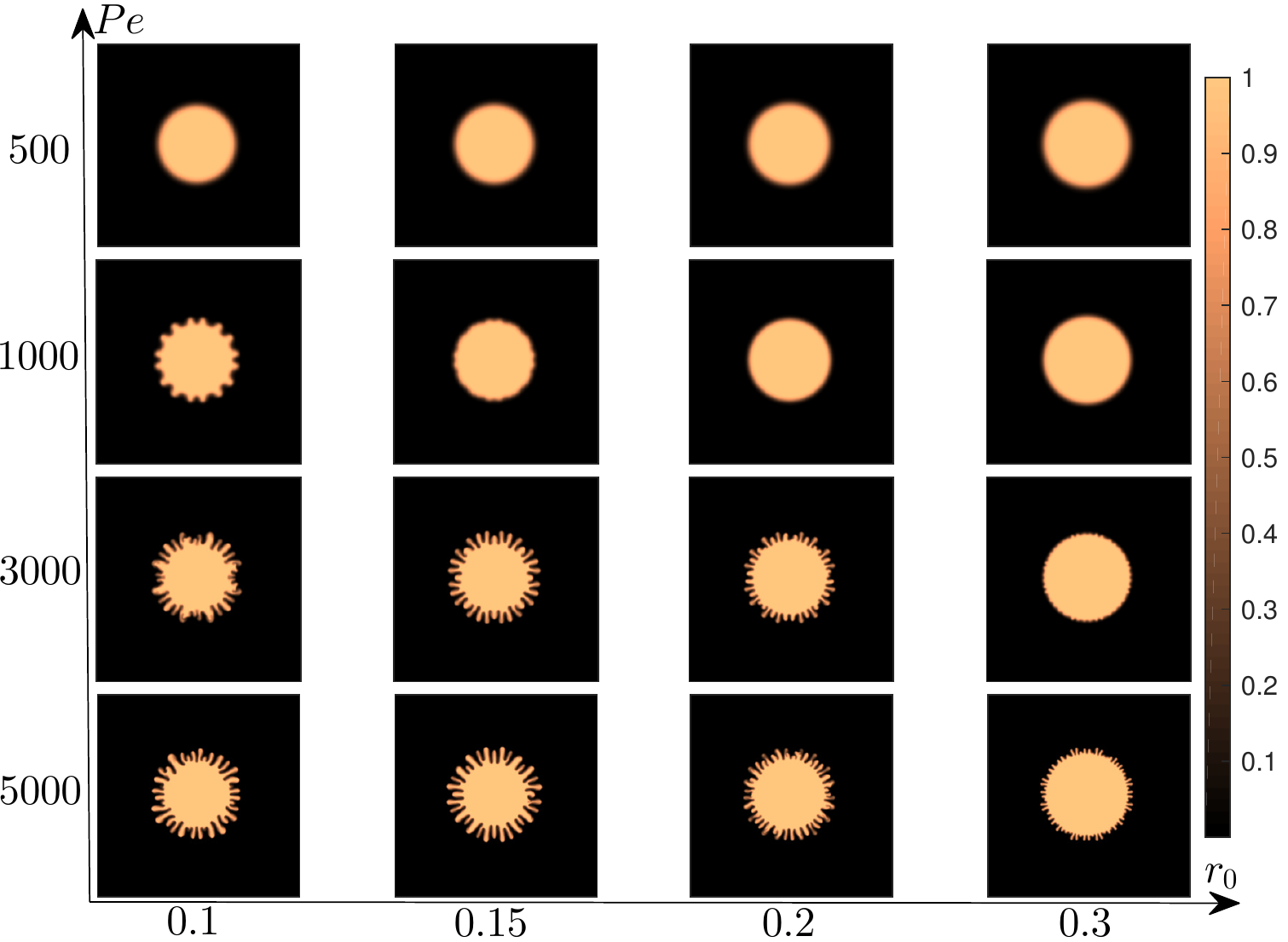}
  \caption{Density plots of concentration at $t=1$ for $(a)$ $Pe=1000$ and various $M, r_0$ obtained numerically. The suppression of instability with increase in $\r$ is clearly evident for $M=2,3$. $(b)$ $M= 2$ but various $Pe, \r$. For a given $M, Pe$ , if there is no instability for a given $\r^*$, then the displacement is stable $\forall r >\r^*$, while a stronger instability is observed $\forall r<\r^*$, if instability is triggered for $\r^*$.}
\label{fig:M_r0}
\end{figure}
\begin{figure}[h]
 \centering
\centerline{\includegraphics[trim={0 0 0 0},clip, scale=0.5]{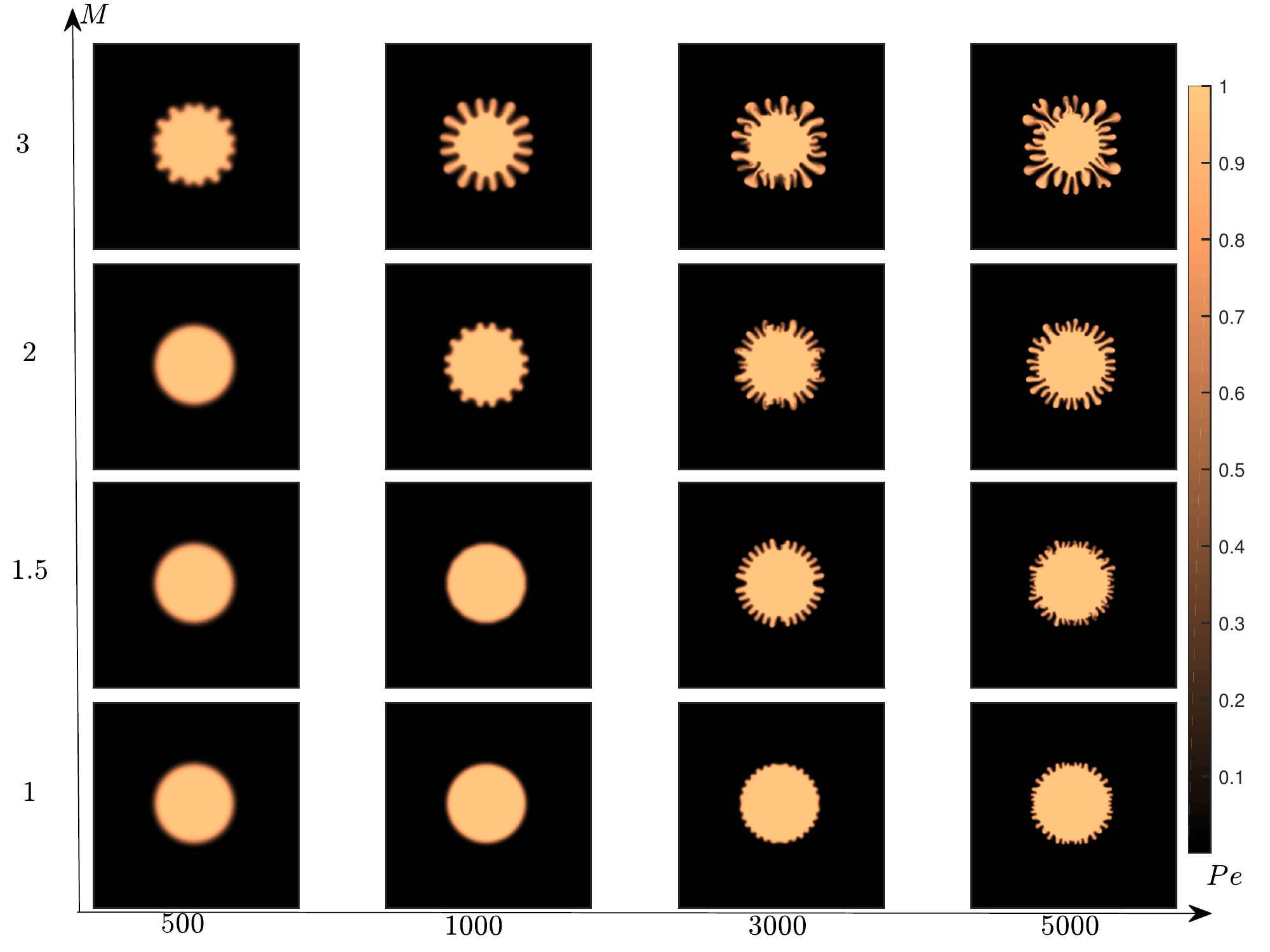}}
  \caption{Density plots of concentration at $t=1$ for $r_0=0.1$ and various $M, Pe$ obtained numerically. The transition in stability on changing either $M$ or $Pe$ is visible. Clearly the set $ \{(2,500),(1.5,500),(1,1000)\} $ of $(M,Pe)$ pair lie in the stable region of the $M-Pe$ parameter space for $\r=0.1$.}
\label{fig:M_PE}
\end{figure}
 
\section{Experiments}
\subsection{Set-up}
\label{App:Expe_set}
We use a radial Hele-Shaw cell with $300 \times 300 \times 10$ mm$^3 $ glass plates and $b = 0.5$ mm gap width. A T-junction made up of a hypodermic syringe needle ($1.2$ mm diameter and $40$ mm length) bent in an L-shape and carefully embedded into a pipe, is used for filling the two fluids in the Hele-Shaw cell. The t-junction avoids the hassle of (a) changing the pipes for different fluids, and (b) having a hole in each glass plate. A syringe pump (Cole-Parmer-D201253) is used for injecting the less viscous fluid. The dynamics are captured with Sony FDR-AX40 camera. 
\begin{figure}[h]
 \centering
\includegraphics[trim={0 0 0 0},clip, scale=0.35]{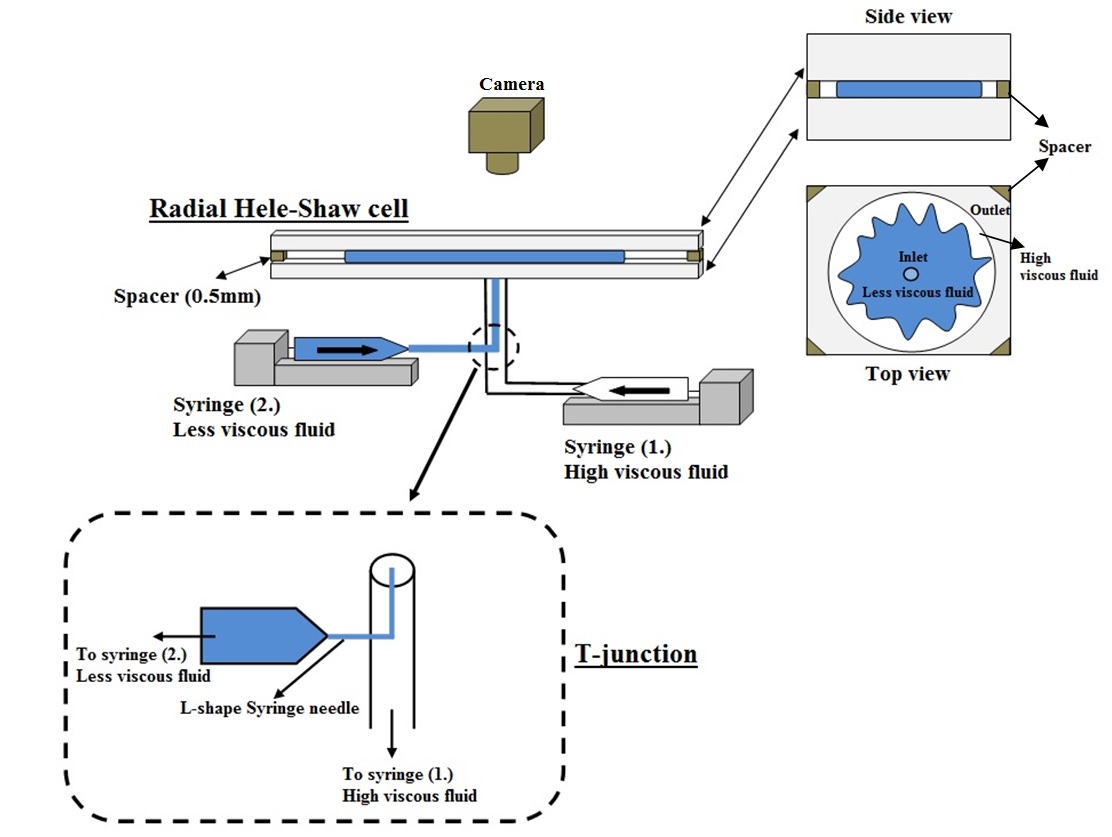}
  \caption{Schematic of the experimental set up, showing the T-junction.}
\label{fig:schematic_exp}
\end{figure}
\subsection{Initial condition}
\label{App:Expe_IC}
The less viscous fluid is injected at a constant flow rate $Q_1$ ml/s for a period of $t_1 $ s in the Hele-Shaw cell initially filled with the more viscous fluid. The initial volume of the injected less viscous fluid, $Q_1 t_1$ ml, is so chosen that it leads to a stable displacement of the more viscous fluid until the invading fluid occupies a circular region of radius $\rdim \;$ mm. Recall that during this stable displacement the interface between the two fluids experiences diffusive spreading proportional to a length $\sqrt{D t_1}$, which also contributes in $\rdim$; we measure $\rdim = \sqrt{D t_1} + \sqrt{Q_1 t_1 /( \pi b)}$. Here, $D = 10^{-9} $ m$^2$/s is the molecular diffusion coefficient of glycerin in water \cite{DErrico2004}. 
\subsection{Effect of increasing $\rdim$ on VF}
\label{App:Expe_r0}
\begin{figure}[h]
 \includegraphics[trim={0 0 0 0},clip, scale=0.15]{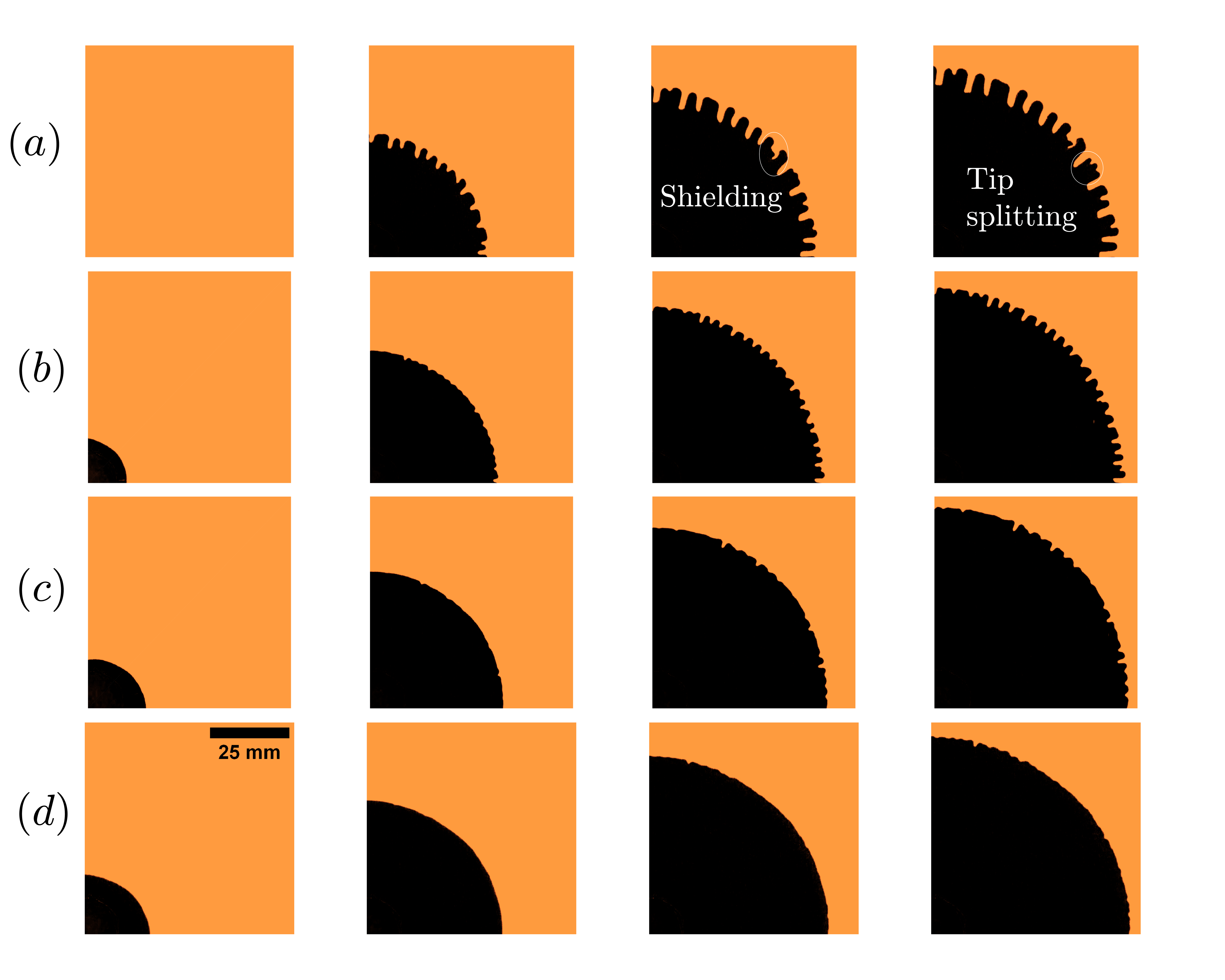}
  \caption{Experimental images at $t= 0, 4800, 8000, 9600 $ ms for $ \rdim = (a)~ 0, (b)~ 15, (c) ~18, (d)~ 20 $ mm.  Tip-splitting and shielding is observable for $\rdim =0$ and are significantly absent for $\rdim \neq 0$. An abated instability with an increase in $\rdim$ is evident. A delay in instability is clearly visible on comparison of the snapshots for different $\rdim$. For visualisation purpose, only one quarter of the experimental images is shown.}
\label{fig:exp_zoomed1}
\end{figure}

\end{document}